\documentclass[aps,prb,twocolumn,superscriptaddress,showpacs,floatfix]{revtex4-1}
\usepackage[dvips]{graphicx}
\usepackage[intlimits]{amsmath}
\usepackage{amssymb}
\DeclareMathOperator{\sgn}{sgn}

\begin{document}
\title{Topological phase transition induced by Zeeman field in topological insulator/superconductor heterostructure}

\author{R.S. Akzyanov}
\affiliation{Moscow Institute of Physics and Technology, Dolgoprudny,
Moscow Region, 141700 Russia}
\affiliation{Institute for Theoretical and Applied Electrodynamics, Russian
Academy of Sciences, Moscow, 125412 Russia}
\affiliation{Dukhov Research Institute of Automatics, Moscow, 127055 Russia}

\author{A.D. Zabolotskiy}
\affiliation{Dukhov Research Institute of Automatics, Moscow, 127055 Russia}

\begin{abstract}
We investigate a topological insulator -- superconductor heterostructure with an Abrikosov vortex in a magnetic field. Large Zeeman field causes a topological phase transition in the system. We found that a Majorana fermion exists at the vortex core in the phase with zero first Chern number and does not exist at the phase with non-trivial first Chern number. This result is supported by the index theorem. This system provides an example in which a Majorana fermion emerges in a system with zero first Chern number.
\end{abstract}


\maketitle

\section{Introduction}
Topological phases are of great interest in modern physics~\cite{ando_rev,bernevig_rev,qi_zhang}. All topological states of matter can be classified by their global symmetries: particle-hole conjugation, time-reversal and their combination -- the chiral symmetry. In each symmetry class the material is characterized by a topological index which can be either trivial or not. If the Hamiltonian preserves the symmetry, then the topological index can be changed only when the gap in the spectrum closes~\cite{schnyder}. This process of changing of topological index with a gap closure is called topological phase transition~\cite{qi_zhang}.


Topological phase transitions in superconductors are of special interest in connection with Majorana fermions~\cite{qi_zhang,Leijnse,Franz}. Such quasiparticles obey non-abelian statistics and may be used for fault-tolerant quantum computations. Usually, Majorana fermions are pinned to the defects in topological superconductors, such as vortices~\cite{Volovik_vortex,fu_kane_device}, hedgehogs~\cite{kane_hedge}, edges~\cite{stone_edge,Alicea} and other boundaries~\cite{kane_defects}. Presence of Majorana fermions is associated with corresponding topological index~\cite{kane_defects}.

One of the most widely used $Z_2$ topological indices for describing presence of Majorana fermions in the system is the evenness of the first Chern number. The first Chern number is the flow of the Berry connection through the Brillouin zone~\cite{schnyder}. For 1D systems the evenness of the first Chern number is equivalent to other topological indices, like Pfaffian invariant~\cite{ghosh} and Uhlmann number~\cite{uhlmann1d}. So, the first Chern number gives full information about the presence of Majorana fermions in 1D systems.

In 2D systems situation differs. It is known from the index theorem~\cite{index_theorem} and from direct calculations~\cite{Volovik_vortex,me} that the presence of the Majorana fermions at the vortex cores depends on the evenness of vorticity. However, the first Chern number is a gauge invariant quantity and does not contain information about vorticity. In Ref.~\citenum{kane_defects} Teo and Kane proposed another topological index for describing Majorana fermions at the defects in class $D$ that is the integral of the Chern-Simons form over the Brillouin zone. For 2D it is equal to the product of the first Chern number and the vorticity mod 2. So, they suggest that even vorticity and even first Chern number are necessary and sufficient for the presence of a Majorana fermion at the vortex core in two dimensions.

So far, the first Chern number shows itself as a manifest of Majorana fermions in superconducting systems~\cite{zhang_qah_sc,roy}. However,
recent works show that separate Majorana fermions can emerge in the non-equilibrium system even if the first Chern number is zero~\cite{zero_chern}. In this paper we will show that Majorana fermion can emerge at the vortex core in the equilibrium system with zero first Chern number.

We study a topological insulator/$s$-wave superconductor heterostructure with a vortex in a magnetic field. Large Zeeman field causes topological phase transition in the system. The first Chern number is trivial in the phase with small Zeeman field and non-trivial in the phase with large Zeeman field. Index theorem and direct solution of BdG equations show that a Majorana fermion emerges at the vortex in the phase with small Zeeman field, where the first Chern number is zero. That is in contradiction with known results for the topological indices. Possible experimental realization of such topological phase transition is discussed.

\section{Calculating topological indices}
In this section we study topological indices of the topological insulator/superconductor (TI/SC) heterostructure. Singlet superconductivity is induced by proximity effect at the metallic surface states of topological insulator. The resulting BdG Hamiltonian in the presence of Abrikosov vortex can be written in Nambu basis $(u_{\uparrow},u_{\downarrow},v_{\downarrow},-v_{\uparrow})^T$ in units $\hbar=1$ as
\begin{eqnarray}\label{ham}
H= v (\sigma_x k_x + \sigma_y k_y)\tau_z + \Delta \tau_x e^{il\theta\tau_z}-U\tau_z-({\bf B}\cdot\sigma)
\end{eqnarray}
where $\sigma_i$ acts in spin space, $\tau_i$ acts in the particle-hole space, $p_i$ is the momentum, $v$ is the Fermi velocity of the surface states, $U$ is the shift from the Dirac point, $B=(B_x,B_y,B_z)$ is the Zeeman field, axis $z$ is perpendicular to the surface,  $\Delta$ is the modulus of superconducting order parameter, $\theta$ is the phase of the order parameter and $l$ is the vorticity. This Hamiltonian has unitary particle-hole symmetry $\Xi^2=+1$ and broken time-reversal symmetry:
\begin{equation}\label{symmetries}
\begin{array}{lcl}
\Xi H(k) \Xi = - H(-k)& \quad &\Xi=\sigma_y\tau_yK 
\\
T H(k) T \neq + H(-k)& \quad &T=i\sigma_yK
\end{array}
\end{equation}
where $\Xi$ is the particle-hole conjugation, $T$ is the time-reversal, $K$ means complex conjugation.
These global symmetries define symmetry class $D$. In that class the presence of Majorana fermions at the point defects is associated with a $Z_2$ index.

\begin{figure}[ht]
\begin{center}
\resizebox{0.6\columnwidth}{!}{\includegraphics{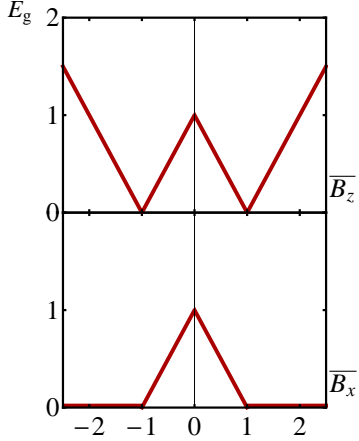}}
\end{center}
\caption{\label{fig:gap} Gap in the spectrum of the Hamiltonian \eqref{ham} versus magnetic field in dimensionless units: $\overline{B_i}=B_i/\sqrt{\Delta^2+U^2}$,  $\overline{E_\textrm{g}}=E_\textrm{g}/\sqrt{\Delta^2+U^2}$. Upper graph is plotted versus purely out-plane magnetic field, lower graph is plotted versus purely in-plane magnetic field.}
\end{figure}

We begin our investigation with spectrum analysis. Magnetic field closes the gap $E_\textrm{g}$ at $|{\bf B}|=\sqrt{\Delta^2+U^2}$ and re-opens the gap at larger values of the magnetic field (Fig.~\ref{fig:gap}). Note that the value of the re-opened gap is proportional to the out-of-plane component of magnetic field, while in-plane magnetic field only closes superconducting gap and does not re-open it.

 We have calculated the first Chern number $C_1$
\begin{eqnarray}\label{c1}
C_1=\frac{1}{2\pi} \sum \limits_{E_n<0}\int \limits
 d^2k \ (\partial_x A_y^{nn}- \partial_y A_x^{nn})
\end{eqnarray}
where $A_j^{lm}=-i\langle u_l|\frac{\partial}{\partial k_j} u_m\rangle$ is the Berry connection, $u_n$ is the Bloch vector of the band $n$.
Direct calculation of the first Chern number \eqref{c1} of the Hamiltonian given by \eqref{ham} shows that
\begin{eqnarray}\label{rll}
C_1=0, \quad &|{\bf B}|<\sqrt{\Delta^2+U^2},& \quad \Delta \neq 0 \nonumber \\
C_1= -\sgn B_z, \quad &|{\bf B}|>\sqrt{\Delta^2+U^2},& \quad B_z\neq 0
\end{eqnarray}
This is not a surprise, since the regime $|{\bf B}|<\sqrt{\Delta^2+U^2}$ is adiabatically connected to the time-reversal state with $|\mathbf{B}|=0,\ l=0$, where the first Chern number has to be zero. In the case of large magnetic field $|{\bf B}|>\sqrt{\Delta^2+U^2}$, the system is adiabatically connected to the state with $\Delta=0$ and unity first Chern number, that exhibits anomalous quantum hall effect~\cite{qah_zhang}.

In the case of 2D topological insulator in class $D$, the $Z_2$ index can be expressed through Chern-Simons form:
\begin{eqnarray}
Z_2^{C}=\int
 \mathcal{Q}_3\,\mathrm{mod}\,2\\
\mathcal{Q}_3=\mathrm{Tr}(\mathcal{A}\wedge d\mathcal{A} + \frac 23 \mathcal{A}\wedge\mathcal{A}\wedge\mathcal{A}) \label{q3}
\end{eqnarray}
where integration is performed over the two-dimensional momentum space and the one-dimensional space of directions around the defect in real space,
$\wedge$ is the wedge product of the differential forms. $\mathcal{A}$ is the Berry connection $2\times2$ matrix (since the band indices run only through the filled bands) with elements being differential 1-forms defined as $\mathcal{A}^{lm}=\sum A^{lm}_kdp_k$, where $p_k$ runs through $k_1$, $k_2$, and $\theta$. The wedge products being applied to matrices of differential forms acts as a usual matrix commutator with outer product of their elements. Thus, the argument of the trace in \eqref{q3} is a $2\times2$ matrix with differential 3-forms, proportional to $dk_1\wedge dk_2\wedge d\theta$, being its elements.

It has been shown \cite{kane_defects} that this $Z_2$ index can be expressed through the first Chern number by the formula
\begin{eqnarray}
\label{Z2C}
Z_2^{C}= C_1 \times l \,\, \mathrm{mod} \, 2
\end{eqnarray}
where $C_1$ is defined by the Eq.~\eqref{c1}. So, $Z_2^C=1$ is nontrivial only if magnetic field is  large $|{\bf B}|>\sqrt{\Delta^2+U^2},\, B_z\neq 0$ and vorticity is odd $l \,\mathrm{mod}\,2 = 1$.

Luckily, for TI/SC heterostructure we can independently compare the answer given by Chern number analysis with the rigorous index theorem.
To perform that comparison,
we apply Atyah-Zinger type index theorem for TI/SC in the same way as it was done in Ref.~\citenum{index_theorem}. That theorem states that number of zero-modes is equal to the index of the Dirac operator.
In case of $U=|\mathbf{B}|=0$ the BdG Hamiltonian can be written using Dirac matrices
\begin{eqnarray}
H=\sum\limits_{a=1,2} v\gamma_a i\nabla_a + \Gamma_a n_a
\end{eqnarray}
where $\gamma_1=\sigma_x\tau_z$, $\gamma_2=\sigma_y\tau_z$, $\Gamma_1=\tau_x$, $\Gamma_2=\tau_y$ are Dirac matrices with anticommutation relations $\{\gamma_i, \gamma_j\}=\{\Gamma_i, \Gamma_j\}=\delta_{ij}$, $\{\Gamma_i,\gamma_j\}=0$; $\mathbf{n}=(\cos l\theta,-\sin l\theta)\Delta$ 
describes complex order parameter and $\nabla = (\partial_x,\partial_y)$. Chiral operator $\gamma_5$ can be written as a product of all gamma matrices $\gamma_5=-\gamma_1\gamma_2\Gamma_1\Gamma_2$. Index theorem implies that number of zero modes of the Hamiltonian is the winding number of the scalar field
\begin{eqnarray}\label{index_theorem}
\mathrm{ind}\, H = -\frac 1{2\pi} \int d^i x\,\epsilon^{ab} \hat{n}_a \partial_i \hat{n}_b
\end{eqnarray}
where $\hat{n}={\bf n}/|{\bf n}|$. Thus, the right side of the equation is just the sum of all vorticities. So, $Z_2$ index that decribes presence of odd number of Majorana fermions in the vortex is equal to the oddness of vorticity $Z_2^{\textrm{ind}}=l\, \mathrm{mod}\, 2$.

This index theorem for TI/SC is generalized for non-zero $B_i$ and $U$ that do not close the gap~\cite{index_theorem_extended}. Such terms violate chiral symmetry. However, these terms do not change the value of $Z_2^{\textrm{ind}}$. Thus, $Z_2^\textrm{ind}$ index that is valid for all $U$ and ${\bf B}$ that do not close the gap from the state with $U=|\mathbf{B}|=0$ is equal to the oddness of vorticity
\begin{eqnarray}\label{z2ind}
Z_2^{\textrm{ind}}=l \,\mathrm{mod}\, 2, \quad |{\bf B}|<\sqrt{\Delta^2+U^2}.
\end{eqnarray}

 As we can see, the result, given by the index theorem, does not contain information about the first Chern number. Moreover, this result is in contradiction with Eq.~\eqref{Z2C} since $Z_2^{C} \neq Z_2^\textrm{ind}$ for odd vorticities $l$.

 To find out which topological number provides correct description of the presence of Majorana fermion at the vortex core we analytically solve BdG equations for the vortex in a Zeeman field that is perpendicular to the surface ${\bf B}=(0,0,B_z)$ with unit vorticity $l=1$.

 We can define a transformation of the basis:
\begin{eqnarray}\label{symm}
\nonumber
\psi &=& \exp[-i(\tau_z-\sigma_z)/2 +i\mu\theta]F^{\mu}(r), \\
F^{\mu} &=& (f^{\mu}_1,f^{\mu}_2,f^{\mu}_3,-f^{\mu}_4)^T
\end{eqnarray}
where $\mu$ is the total angular momentum of the state. Majorana fermion is the solution with zero energy
that must be its own antiparticle $\psi=\Xi \psi$, where $\Xi$ is defined by \eqref{symmetries}. This condition implies that  Majorana fermion must have zero angular momentum $\mu=0$. After transformation of the Hamiltonian~\eqref{ham} using symmetry transformation~\eqref{symm} for $\mu=0$ we obtain that
\begin{eqnarray}\label{final}
\nonumber
i{v}\!\left (\frac d{dr}\!+\!\frac {1}{r}\!\right)\!f_2\!+\!{\Delta}f_3\!-\!(U\!+\!B_z)f_1\!\!=\!0,\quad
\\ \nonumber
i{v}\!\frac d{dr}\!f_1\!-\!{\Delta}f_4\!-\!(U\!-\!B_z)f_2\!\!=\!0,\quad
\\
i{v}\frac d{dr}\!f_4\!+\!{\Delta}f_1\!+\!(U\!-\!B_z)f_3\!\!=\!0,\quad
\\
\nonumber
i{v}\!\left (\frac d{dr}\!-\!\frac {1}{r}\!\right)\!f_3\!-\!{\Delta}f_2\!+\!(U\!+\!B_z)f_4\!\!=\!0.\quad
\end{eqnarray}
 After the following substitution
\begin{eqnarray}\label{Center}
 \nonumber
 X_1=if_1+f_4,\quad X_2=if_1-f_4,  \\
 Y_1=if_2-f_3,\quad Y_2=if_2+f_3,
\end{eqnarray}
the system of equations~\eqref{final}  decouples into the two subsystems
\begin{eqnarray}
{v}\!\left (\frac d{dr}\!-\!\frac {\Delta}{v}\!\right)\!X_1\!-i\!\left(B_z\!-\!U\right)Y_1\!\!=\!0, \nonumber \\
{v}\!\left (\frac d{dr}\!+\frac {1}{r}-\!\frac {\Delta}{v}\right)\!Y_1\!
+i\!\left(B_z\!+\!U\right)X_1\!\!=\!0,
\end{eqnarray}
and
\begin{eqnarray}
{v}\!\left (\frac d{dr}\!+\!\frac {\Delta}{v}\!\right)\!X_2\!-i\!\left(B_z\!-\!U\right)Y_2\!\!=\!0, \nonumber \\
{v}\!\left (\frac d{dr}\!+\frac {1}{r}+\!\frac {\Delta}{v}\right)\!Y_2\!
+i\!\left(B_z\!+\!U\right)X_2\!\!=\!0.
\end{eqnarray}
These two subsystems differ only by the sign at $\Delta$. Thus, after the transformation we get $X(Y)_{\alpha}=\tilde{X}(\tilde{Y})_{\alpha}\times e^{\lambda\int \limits_0^r \Delta(r')/v\,dr'}$, where $\lambda=1$ for the first subsystem $\alpha=1$ and $\lambda=-1$ for the second subsystem $\alpha=2$, solution of these two systems can be rewritten in terms of solution of the following equation
\begin{eqnarray}\label{bes}
i\left(B_z\!-\!U\right)\tilde{Y}_{\alpha}\!=\!v\frac {d}{dr}\tilde{X}_{\alpha},\nonumber\\
\frac {d^2}{dr^2} \tilde{X}_{\alpha}\!+\!\frac 1r \frac {d}{dr} \tilde{X}_{\alpha}\!-\!\left(B_z^2-U^2\right)\tilde{X}_{\alpha}\!=\!0
\end{eqnarray}
Solutions of the Eq.~\eqref{bes} can be expressed in terms of Bessel functions of either real or complex argument, depending on the sgn$(U^2-B_z^2)$. Solutions integrable at $r=0$ can be expressed in terms of the Bessel function of the first kind with zero order
\begin{eqnarray}
\tilde{X}_{\alpha}=J_0\left(\frac rv \sqrt{U^2-B_z^2}\right)
\end{eqnarray}
Solution of the first subsystem $\alpha=1$ that is $X_1=J_0\left(\frac rv \sqrt{U^2-B_z^2}\right)\times e^{\int \limits_0^r \Delta(r')/v\,dr'}$ is not integrable at any values of $(\Delta, U, B_z)$, so $X_1=Y_1=0$. Solution of the second subsystem $\alpha=2$ that is $X_2=J_0\left(\frac rv \sqrt{U^2-B_z^2}\right)\times e^{-\int \limits_0^r \Delta(r')/v\,dr'}$ is integrable if $|U|\geq |B_z|$ for any nonzero $\Delta$. If $|U| < |B_z|$ then the solution for large argument can be expressed as $X_2 \sim e^{-\Delta/v} e^{\sqrt{B_z^2-U^2}}$, which is integrable only if $\Delta > \sqrt{B_z^2-U^2}$. So, the condition for the Majorana fermion localized at the vortex core can be written as follows:
\begin{eqnarray}
\Delta^2 + U^2 - B_z^2 > 0
\end{eqnarray}
As we can see, that condition is in contradiction with the answer given by Chern number analysis, see Eq.~\eqref{
rll} and in agreement with index theorem, see Eq.~\eqref{
z2ind}.
 Majorana spinor can be expressed as follows for $|U|\geq |B_z|$
\begin{eqnarray}
\label{edge_m_1}
\psi=C e^{-i\frac{\pi}{4}\sigma_z}  e^{-\int
\limits_0^r \frac{\Delta(r')}{v} dr'}
\begin{bmatrix}
J_0\left(\frac rv \sqrt{U^2-B_z^2}\right)\\
J_1\left(\frac rv \sqrt{U^2-B_z^2}\right)e^{-i\theta}\\
J_1\left(\frac rv \sqrt{U^2-B_z^2}\right)e^{i\theta}\\
-J_0\left(\frac rv \sqrt{U^2-B_z^2}\right)
\end{bmatrix}.
\end{eqnarray}
For $|U|< |B_z|$ Majorana fermion solution can be expressed as
\begin{eqnarray}
\psi=C e^{-i\frac{\pi}{4}\sigma_z}  e^{-\int
\limits_0^r \frac{\Delta(r')}{v} dr'}
\begin{bmatrix}
I_0\left(\frac rv \sqrt{B_z^2-U^2}\right)\\
I_1\left(\frac rv \sqrt{B_z^2-U^2}\right)e^{-i\theta}\\
I_1\left(\frac rv \sqrt{B_z^2-U^2}\right)e^{i\theta}\\
-I_0\left(\frac rv \sqrt{B_z^2-U^2}\right)
\end{bmatrix}.
\end{eqnarray}
where $I_m(x)$ is modified Bessel function of the first kind. It is easy to check that $\Xi \psi = \psi$, hence $\psi$ is Majorana fermion.
We note that for $|B_z| \geq |U|$ there is no spatial oscillation of the wavefunction in contrast to the opposite case $|B_z| < |U|$.

\section{Discussion}
\begin{figure}[ht]
\begin{center}
\resizebox{0.8\columnwidth}{!}{\includegraphics{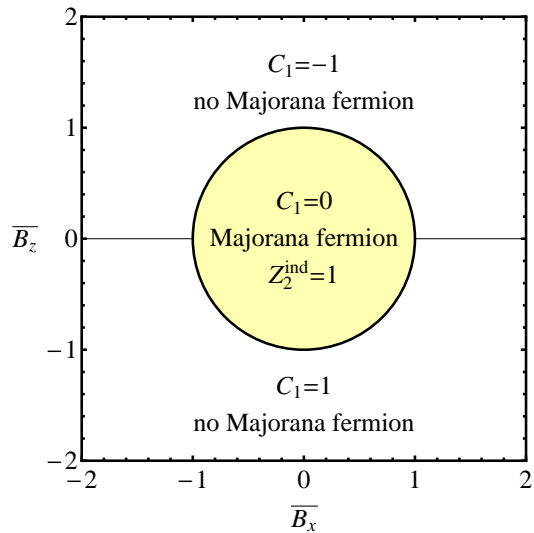}}
\end{center}
\caption{\label{fig:phase} Topological phase diagram of the topological insulator/superconductor structure in out-plane $\overline{B_z}=B_z/\sqrt{\Delta^2+U^2}$ and in-plane $\overline{B_x}=B_x/\sqrt{\Delta^2+U^2}$ magnetic fields. $C_1$ is the first Chern number. $Z_2^{\textrm{ind}}$ is the winding given by the index theorem. The filled circle represents the area with the Majorana fermion at the vortex core. }
\end{figure}

We plot the topological phase diagram of the topological insulator/superconductor heterostructure in a magnetic field (Fig. \ref{fig:phase}) .

In this figure we can see that a Majorana fermion exists at the vortex core in the phase with small magnetic field $|{\bf B}|<\sqrt{\Delta^2+U^2}$, where the first Chern number is zero. At larger magnetic field $|{\bf B}|>\sqrt{\Delta^2+U^2}$ first Chern number is non-trivial, however, there is no Majorana fermion at the vortex core.
As we can see, the Chern number analysis gives the wrong answer in comparison with both index theorem and strict solving of BdG equations. So, topological index that provides the information about presence of Majorana fermions at the defects in arbitrary systems should not be always proportional to the first Chern number. Why do we need that index?
Additional terms in hamiltonian, such as hexagonal warping, band curvature, anisotropy, etc., make it difficult to solve BdG equations directly. Proper topological index can provide a quick answer: Can a Majorana fermion exist in the system, and what parameters of the system do we need for that?

Land\'{e} factor can be very large in topological insulators~\cite{g_factor}, up to $g\sim 80$. Taking characteristic values of Land\'e factor $g=65$ for Bi$_2$Te$_3$, superconducting order parameter $\Delta=1$ meV for NbTi and considering chemical potential that lies at the Dirac point, then the topological phase transition occurs at magnetic field $B_c=\Delta/g\mu_B \sim 0.23$ T which is smaller than the second critical field of NbTi $B_{c2}=10$ T~\cite{schmidt}. Also, it is known that for the thin film of superconductor $d \ll \lambda_{L}$ the second critical field can be be significantly enhanced $B_{c2}=2\sqrt{6}B_0\lambda_{L}/d$ in comparison with the second critical field for massive superconductor $B_0$~\cite{schmidt}. Recent experiments show that for Pb thin island the second critical field can be as large as 1 T~\cite{Rodichev}. Fine tuning of chemical potential to the Dirac point should be done in order to achieve condition $B>\sqrt{\Delta^2+U^2}$.

What observables can be measured? Re-entrant superconductivity can be observed by the transport or STM measurements. We claim that if Zeeman field is large enough $B>\sqrt{\Delta^2+U^2}$ then the Majorana fermion cannot exist at the vortex core. Consider a cylindrical hole in the superconductor layer placed on the top of the topological insulator. In the presence of small magnetic field that is perpendicular to the surface a vortex with a Majorana fermion can be trapped by the hole. Majorana fermion can be observed as a robust zero-bias peak if $B<\sqrt{\Delta^2+U^2}$. Now we apply in-plane magnetic field $B>\sqrt{\Delta^2+U^2}$ and after re-opening of the gap Majorana fermion peak should disappear.

We have studied a topological insulator/superconductor structure with the vortex in the magnetic field.
This is an example of the system in which a Majorana fermion exists at the vortex core in the phase with zero first Chern number and does not exist in the phase with non-zero first Chern number. That result is supported by the index theorem.

\begin{acknowledgements}
We acknowledge partial support by the Dynasty Foundation and ICFPM (MMK), the Ministry of Education and Science of the Russian Federation Grant No.~14Y26.31.0007, RFBR Grant No.~15-02-02128.
\end{acknowledgements}

\end{document}